\documentclass[12pt]{article}
\usepackage{setspace}
\usepackage{epsfig}
\usepackage{amssymb}
\def\be{\begin{equation}}
\def\ee{\end{equation}}
\def\ba{\begin{array}}
\def\ea{\end{array}}
\def\beqn{\begin{eqnarray}}
\def\eeqn{\end{eqnarray}}

\def\bt{\begin{tabular}}
\def\et{\end{tabular}}
\def\bc{\begin{center}}
\def\ec{\end{center}}

\def\vckm{$V_{{\rm CKM}}$}

\def\sin2{sin$2\beta$}

\begin{document}

\title{Possible textures of the fermion mass matrices}

\author{Manmohan Gupta, Gulsheen Ahuja\\
\\
{\it Department of Physics, Centre of Advanced Study, P.U.,
Chandigarh, India.}\\ {\it Email: mmgupta@pu.ac.in}}

\maketitle

\begin{abstract}
Texture specific fermion mass matrices have played an important
role in understanding several features of fermion masses and
mixings. In the present work, we have given an overview of all
possible cases of Fritzsch-like as well as non Fritzsch-like
texture 6 and 5 zero fermion mass matrices. Further, for the case
of texture 4 zero Fritzsch-like quark mass matrices, the issue of
the hierarchy of the elements of the mass matrices and the role of
their phases have been discussed. Furthermore, the case of texture
4 zero Fritzsch-like lepton mass matrices has also been discussed
with an emphasis on the hierarchy of neutrino masses for both
Majorana and Dirac neutrinos.
\end{abstract}

\section{Introduction}
Understanding fermion masses and mixings is one of the outstanding
problem of present day particle physics. The idea of quark mixing
phenomena was initiated by Cabibbo in 1963 \cite{cabibbo},
subsequently generalized to two generations by Glashow,
Illiopoulos, Maiani  \cite{glashow} and finally to three
generations by Kobayashi and Maskawa  \cite{kobayashi}. This has
been tested to a great accuracy and is well accommodated by the
Standard Model (SM). Recently, flavor mixing has also been
observed in the case of neutrinos implying the existence of non
zero, non degenerate neutrino masses necessitating the need to
look beyond SM. Also, one has to go beyond the SM in order to
understand the pattern of quark masses and mixing parameters as in
the SM the quark mass matrices are completely arbitrary. In view
of the relationship of fermion mixing phenomena with that of
fermion mass matrices, the understanding of the above mentioned
issues of flavor physics essentially implies formulating fermion
mass matrices.

While on the one hand, Grand unified Theories (GUTs) have provided
vital clues for understanding the relationship of fermion mass
matrices between quarks and leptons, on the other hand, horizontal
symmetries \cite{horizontal} have given clues for the relationship
between different generation of fermions. Ideas such as extra
dimensions \cite{extradim} have also been invoked to understand
the flavor puzzle. Unfortunately, at present it seems that we do
not have any theoretical framework which provides a viable and
satisfactory description of fermion masses and mixings.

The lack of a convincing fermion flavor theory from the `top down'
perspective necessitates the need for formulating fermion mass
matrices from a `bottom up' approach. The essential idea behind
this approach is that one tries to find the phenomenological
fermion mass matrices which are in tune with the low energy data
and can serve as guiding stone for developing more ambitious
theories. In this context, initially several {\it ans\"{a}tze}
\cite{frzans,ansatze} were suggested for quark mass matrices. One
of the successful {\it ans\"{a}tze} incorporating the ``texture
zero'' approach was initiated by Fritzsch \cite{frzans}. {\it A
particular texture structure is said to be texture $n$ zero, if it
has $n$ number of non-trivial zeros, for example, if the sum of
the number of diagonal zeros and half the number of the
symmetrically placed off diagonal zeros is $n$}.

The detailed plan of the article is as follows. In Section
(\ref{tsmm}), we discuss some of the broad features pertaining to
quark and lepton texture specific mass matrices. The relationships
of the fermion mass matrices and mixing matrices have been
presented in Section (\ref{form}). Present status of the quark and
neutrino mass and mixing parameters have been given in Section
(\ref{inputs}). The details pertaining to texture 6, 5, 4 zero
quark and lepton mass matrices have respectively been presented in
Sections (\ref{quarkmm}) and (\ref{lepmm}). Finally, in Section
(\ref{summ}) we summarize and conclude.

\section{Texture specific mass matrices \label{tsmm}}
\subsection{Quark mass matrices}
The mass matrices, having their origin in the Higg's fermion
couplings, are arbitrary in the SM, therefore the number of free
parameters available with a general mass matrix is larger than the
physical observables. For example, if no restrictions are imposed,
there are $36$ real free parameters in the two $3 \times 3$
general complex mass matrices, $M_U$ and $M_D$, which in the quark
sector need to describe ten physical observables, i.e., six quark
masses, three mixing angles and one CP violating phase. Similarly,
in the leptonic sector, physical observables described by lepton
mass matrices are six lepton masses, three mixing angles and one
CP violating phase for Dirac neutrinos (two additional phases in
case neutrinos are Majorana particles). Therefore, to develop
viable phenomenological fermion mass matrices one has to limit the
number of free parameters in the mass matrices.

In this context, it is well known that in the SM and its
extensions wherein the right handed fields in the Lagrangian are
SU(2) singlets, without loss of generality, the mass matrices can
be considered as hermitian. This immediately brings down the
number of real free parameters from 36 to 18, which however, is
still a large number compared to the number of observables. To
this end, Fritzsch \cite{frzans} initiated the idea of texture
specific mass matrices which on the one hand imparted
predictability to mass matrices while on the other hand, it paved
the way for the phenomenology of texture specific mass matrices.
To define the various texture specific cases, we present the
typical Fritzsch like texture specific hermitian quark mass
matrices, for example,
\be
 M_{U}=\left( \ba{ccc}
0 & A _{U} & 0      \\ A_{U}^{*} & D_{U} &  B_{U}     \\
 0 &     B_{U}^{*}  &  C_{U} \ea \right), \qquad
M_{D}=\left( \ba{ccc} 0 & A _{D} & 0      \\ A_{D}^{*} & D_{D} &
B_{D}     \\
 0 &     B_{D}^{*}  &  C_{D} \ea \right),
\label{nf2zero}\ee
 where $M_{U}$ and $M_{D}$ correspond to up and
down mass matrices respectively. It may be noted that each of the
above matrix is texture 2 zero type with $A_{i}
=|A_i|e^{i\alpha_i}$
 and $B_{i} = |B_i|e^{i\beta_i}$, where $i= U,D$.

The texture 6 zero Fritzsch mass matrices can be obtained from the
above mentioned matrices by taking both $D_{U}$ and $D_{D}$ to be
zero, which reduces the matrices $M_{U}$ and $M_{D}$ each to
texture 3 zero type. This Fritzsch {\it ans\"{a}tze} \cite{frzans}
as well as some other {\it ans\"{a}tze} \cite{ansatze} were ruled
out because of the large value predicted for $|V_{cb}|$ due to the
high `t' quark mass.

Further, a few other texture 6 zero mass matrices were analyzed by
Ramond, Roberts and Ross \cite{rrr} revealing that these matrices
were again ruled out because the predicted value of $|V_{cb}|$
came out to be much larger than the available data at that time.
They also explored the question of connection between
phenomenological quark mass matrices considered at low energies
and the possible mass patterns at the GUT scale and showed that
the texture structure of mass matrices is maintained as we come
down from GUT scale to $m_Z$ scale. This important conclusion also
leads to the fact that the texture zeros of fermion mass matrices
can be considered as phenomenological zeros, thereby implying that
at all energy scales the corresponding matrix elements are
sufficiently suppressed in comparison with their neighboring
counterparts. This, therefore, opens the possibility of
considering lesser number of texture zeros.

Besides Ramond, Roberts and Ross \cite{rrr}, several authors
\cite{fri2000}-\cite{5zero} then tried to explore the texture 5
zero quark mass matrices. Fritzsch-like texture 5 zero matrices
can be obtained by taking either $D_{U}$ = 0 and $D_{D} \neq 0$ or
$D_{U} \neq 0$ and $D_{D}$ = 0 in Eq. (\ref{nf2zero}), thereby
giving rise to two possible cases of texture 5 zero mass matrices
pertaining to either $M_U$ or $M_D$ being texture 3 zero type
while the other being texture 2 zero type. These analyses reveal
that texture 5 zero mass matrices although not ruled out
unambiguously yet are not able to reproduce the entire range of
data. Further, the issue of the phases of the mass matrices,
responsible for CP violation, was not given adequate attention in
these analyses.

As an extension of texture 5 zero mass matrices, several authors
 \cite{fri2000,group4zero}-\cite{cps} carried out the study of
the implications of the Fritzsch-like texture 4 zero mass
matrices. It may be noted that Fritzsch-like texture 4 zero mass
matrices can be obtained by considering both $M_U$ and $M_D$, with
non zero $D_i (i=U,D)$ in Eq. (\ref{nf2zero}), to be texture 2
zero type. Although from the above mentioned analyses one finds
that texture 4 zero mass matrices were able to accommodate the
quark mixing data quite well, however it may be noted that these
analyses assumed `strong hierarchy' of the elements of the mass
matrices as well as explored only their limited domains. Further,
in the absence of any precise information about CP violating phase
${\delta}$ and related parameters, again adequate attention was
not given to the phases of the mass matrices.

Recent refinements in quark mixing data motivated several authors
\cite{hallraisin}-\cite{s2b} to have a re-look at the
compatibility of Fritzsch like texture 4 zero mass matrices with
the quark mixing data. In particular, using assumption of `strong
hierarchy' of the elements of the mass matrix defined as $D_i <
|B_{i}|
< C_i, (i=U,D)$, having its motivation in the
hierarchy of the quark mixing angles several attempts
\cite{hallraisin}-\cite{branco} were made to predict the value of
precisely known parameter sin$\,2\beta$. Unfortunately, the value
of sin$\,2\beta$ predicted by these analyses came out to be in
quite disagreement with its precisely known value. A somewhat
detailed and comprehensive analyses of texture 4 zero quark mass
matrices for the first time was carried out by Xing and Zhang
\cite{xingzhang}, in particular they attempted to find the
parameter space available to the elements of mass matrices. Their
analysis has also given valuable clues about the phase structure
of the mass matrices, in particular for the strong hierarchy case
they conclude that only one of the two phase parameters plays a
dominant role. Subsequently, attempts \cite{cps,s2b} have been
made to update and broaden the scope of the analysis carried out
by Xing and Zhang \cite{xingzhang}, in particular regarding the
structural features of the mass matrices having implications for
the value of parameter sin$\,2\beta$.

\subsection{Lepton mass matrices}

In the leptonic sector, one would like to mention that
 the observation of neutrino oscillations
has added another dimension to the issue of fermion masses and
mixing. In fact, the pattern of neutrino masses and mixings seems
to be vastly different from that of quarks. At present, the
available neutrino oscillation data does not throw any light on
the neutrino mass hierarchy, which may be normal/ inverted and may
even be degenerate. Further, the situation becomes complicated
when one realizes that neutrino masses are much smaller than
charged fermion masses as well as it is not clear whether
neutrinos are Dirac or Majorana particles. The situation becomes
more complicated in case one has to understand the quark and
neutrino mixing phenomena in a unified manner.

In this context, to understand the pattern of neutrino masses and
mixings, texture zero approach has also been tried with good deal
of success \cite{framp}-\cite{leptex}. An early attempt to
formulate lepton mass matrices was carried out by Frampton,
Glashow and Marfatia \cite{framp}, wherein assuming a complex
symmetric Majorana mass matrix and considering seven possible
texture 2 zero cases, they carried out the implications of these
for the neutrino oscillation data. Thereafter, several attempts
were made using texture specific lepton mass matrics to explain
the pattern of neutrino masses and mixings. In particular, for
normal hierarchy of neutrino masses, Fukugita, Tanimoto and
Yanagida \cite{fuku} carried out an analysis of Fritzsch-like
texture 6 zero mass matrices. Similarly, Zhou and Xing \cite{zhou}
also carried out a systematic analysis of all possible texture 6
zero mass matrices for Majorana neutrinos with an emphasis on
normal hierarchy of neutrino masses. Recently,
\cite{ourneut6zero,ourneut4zero} for all possible hierarchies of
neutrino masses, for both Majorana as well as Dirac neutrinos,
detailed analyses of Fritzsch-like texture 6, 5 and 4 zero mass
matrices was carried out.

From the above discussion, one finds that texture specific mass
matrices are able to accommodate the quark as well as neutrino
mixing data. This brings into fore the issue of quark-lepton
unification, advocated by Smirnov \cite{qlepuni}, by considering
similar structures for quark and lepton mass matrices. Keeping
this in mind as well as in view of absence of any theoretical
justification for Fritzsch-like mass matrices, recent attempts
\cite{neeluijmpa,neelu6zero} were made to consider non
Fritzsch-like mass matrices for quarks as well as neutrinos.

\section{Relationship of fermion mass matrices and mixing matrices \label{form}}
\subsection{Quark mass matrices and mixing matrix}

In the SM, the quark mass terms for three generations
 of quarks can be expressed as
\be
{\overline q}_{U_L} M_U ~ {q}_{U_R}  + {\overline q}_{D_L} M_D ~
{q}_{D_R}\,,
 \label{mc} \ee
where ${q}_{U_{L(R)}}$ and  ${q}_{D_{L(R)}}$ are the left-handed
(right-handed)
 quark fields for the  up sector
$(u,c,t)$ and down sector $(d,s,b)$ respectively. $M_U$ and $M_D$
are the mass matrices for the up and the down sector  of quarks.
 In order to re-express above equation in terms of the physical quark
fields, one can diagonalize the mass matrices by the following
bi-unitary transformations \be V_{U_L}^{\dagger}M_U V_{U_R} =
M_U^{diag} \equiv {\rm Diag}\, (m_u,m_c,m_t)\,, \label{v11} \ee
\be V_{D_L}^{\dagger}M_D V_{D_R} = M_D^{diag} \equiv {\rm Diag}
\,(m_d,m_s,m_b)\,, \label{v21} \ee
 where $M_{U,D}^{diag}$ are real and diagonal, while $V_{U_L}$ and
  $V_{U_R}$ etc. are complex unitary matrices. The quantities $m_u, m_d$ etc. denote the
  eigenvalues of the mass matrices, i.e. the physical
quark masses. Using Eqs. (\ref{v11}) and (\ref{v21}), one can
rewrite (\ref{mc}) as
 \be  \overline{q}_{U_L}V_{U_L}M_U^{diag}
V_{U_R}^{\dagger} q_{U_R}  + \overline{q}_{D_L} V_{D_L} M_D^{diag}
V_{D_R}^{\dagger} q_{D_R}\, , \ee which can be re-expressed in
terms of physical quark fields as
\be
 \overline{q}^{phys}_{{U}_L} M_U^{diag}
q^{phys}_{U_R}  + \overline{q}^{phys}_{{D}_L}M_D^{diag}
 q^{phys}_{D_R}\, ,\ee
where $q^{phys}_{U_L}\,=\,V_{U_L}^{\dagger} q_{U_L}$ and
 $q^{phys}_{D_L}\,=\,V_{D_L}^{\dagger} q_{D_L}$ and so on.

The mismatch of diagonalizations of up and down quark mass
matrices leads to the quark mixing matrix \vckm, referred to as
the Cabibbo-Kobayashi-Maskawa (CKM) \cite{cabibbo,kobayashi}
matrix given as \be V_{\rm CKM} =  V_{U_L}^{\dagger} V_{D_L}.
 \label{1mix}
\ee The CKM matrix expresses the relationship between quark mass
eigenstates $d,\,s,\,b$ which participate in the strong q$-$q and
q$-\overline{\rm{q}}$ interactions and the interaction eigenstates
or flavor eigenstates $d', s', b'$ which participate in the weak
interactions and are the linear combinations of mass eigenstates,
for example, \be  \left( \ba{c} d'
\\ s'
\\ b' \ea \right) = \left( \ba {lll}
V_{ud} & V_{us} & V_{ub} \\ V_{cd} & V_{cs} & V_{cb} \\ V_{td} &
V_{ts} & V_{tb} \\ \ea \right) \left( \ba {c} d\\ s \\ b \ea
\right), \label {1vckm}\ee where $V_{ud}$, $V_{us}$ etc. describe
the transition of $u$ to $d$, $u$ to $s$ respectively, and so on.

In view of the relationship of the mixing matrix with the mass
matrix, a knowledge of the \vckm elements would have important
implications for the mass matrices. The \vckm, by definition, is a
unitary matrix, hence can be expressed in terms of
 three real angles and six phases. Out of
the six phases, five can be re-absorbed into the quark fields,
therefore, one is left with only one non-trivial phase which is
responsible for CP violation in the SM. There are several
parameterizations of the SM, however the most commonly used
parameterization is the standard parameterization given by
Particle Data Group (PDG) \cite{pdg}.
 The PDG representation of the \vckm is given as
\be V_{\rm CKM} =
 \left( \ba {ccc} c_{12} c_{13} & s_{12} c_{13} &
  s_{13}e^{-i{ \delta}} \\
  -s_{12} c_{23} - c_{12} s_{23} s_{13}e^{i{\delta}} &
 c_{12} c_{23} - s_{12} s_{23}s_{13}e^{i{\delta}}
  & s_{23} c_{13} \\
  s_{12} s_{23} - c_{12} c_{23} s_{13}e^{i{\delta}} &
  - c_{12} s_{23} - s_{12}c_{23} s_{13}e^{i{\delta}} &
  c_{23} c_{13} \ea \right),  \label{1ckm}  \ee
  with $c_{ij}={\rm cos}\,\theta_{ij}$ and
   $s_{ij}={\rm sin}\,\theta_{ij}$.  The angles $\theta_{12}, \theta_{23}$
and $\theta_{13}$ can be chosen to lie in the first quadrant,
whereas the quadrant of ${\delta}$ has physical significance,
therefore cannot be fixed. In the PDG representation, sin${\delta}
\neq 0$ implies the existence of  CP violation. A precise
measurement of \vckm elements and the parameters characterizing
the PDG representation e.g., mixing angles and CP violating phase
$\delta$ will undoubtedly have implications for the mass matrices.

\subsection{Lepton mass matrices and mixing matrix}
The observation of neutrino oscillation phenomenon which
essentially implies the flavor conversion of neutrinos is similar
to the quark mixing phenomenon. This possibility of flavor
conversion was originally examined by B. Pontecorvo and further
generalized by Maki, Nakagawa and Sakata \cite{pmns}.
 The emerging picture that neutrinos are
massive and therefore mix has been proved beyond any doubt and
provides an unambiguous signal of NP.

In the case of neutrinos, the generation of masses is not
straight-forward as they may have either the Dirac masses or the
more general Dirac-Majorana masses. A Dirac mass term can be
generated by the Higgs mechanism with the standard Higgs doublet.
In this case, the neutrino mass term can be written as
\begin{equation}
 \overline{\nu}_{a_{L}} M_{\nu D} {\nu}_{a_{R}} + h.c.,
\end{equation}
where $a$ = $e$, $\mu$, $\tau$ and $\nu_e$, $\nu_\mu$, $\nu_\tau$
are the flavor eigenstates. $M_{\nu D}$ is a complex $3\times 3$
Dirac mass matrix. The mass term mentioned above would also be
characterized by the same symmetry breaking scale such as that of
charged leptons or quarks, therefore, in this case very small
masses of neutrinos would be very unnatural from the theory point
of view. On the other hand, the neutrino might be a Majorana
particle which is defined as is its own antiparticle and is
characterized by only two independent particle states of the same
mass ($\nu^{~}_{\rm L}$ and $\bar{\nu}^{~}_{\rm R}$ or
$\nu^{~}_{\rm R}$ and $\bar{\nu}^{~}_{\rm L}$). A Majorana mass
term, which violates both the law of total lepton number
conservation and that of individual lepton flavor conservation,
can be written either as
\begin{equation}
 \frac{1}{2} \overline{\nu}_{a_{L}} M_L {\nu}^c_{a_R}
+ h.c.~~~~~~~~~{\rm or~as}~~~~~~~~~\frac{1}{2}
\overline{\nu}^c_{a_L} M_R {\nu}_{a_R} + h.c.,
\end{equation}
where $M_l$ and $M_R$ are complex symmetric matrices.

A simple extension of the SM is to include one right handed
neutrino in each of the three lepton families, while the
Lagrangian of the electroweak interactions is kept invariant under
$SU(2)_L \times U(1)_Y$ gauge transformations. This can be shown
to lead to Dirac-Majorana mass terms which further lead to the
famous seesaw mechanism \cite{seesaw} for the generation of small
neutrino masses, e.g.,
 \be M_{\nu}=-M_{\nu D}^T\,(M_R)^{-1}\,M_{\nu D},
\label{seesaweq2} \ee \noindent where $M_{\nu D}$ and $M_R$ are
respectively the Dirac neutrino mass matrix and the right-handed
Majorana neutrino mass matrix.

The seesaw mechanism is based on the assumption that, in addition
to the standard Higgs mechanism of generation of the Dirac mass
term, there exists a beyond the SM mechanism of generation of the
right-handed Majorana mass term, which changes the lepton number
by two and is characterized by a mass $M \gg m$. The Dirac mass
term mixes the left-handed field $\nu_L$, the component of a
doublet, with a single field $(\nu^c)_R$. As a result of this
mixing the neutrino acquires Majorana mass, which is much smaller
than the masses of leptons or quarks.

Similar to the quark sector, the lepton mass matrices can be
diagonalized by bi-unitary transformations and the corresponding
mixing matrix obtained, known as Pontecorvo-Maki-Nakagawa-Sakata
(PMNS) or lepton mixing matrix \cite{pmns}, is given as
\be
V_{\rm PMNS}= V^{\dagger}_{l_{L}} V_{\nu_{L}}. \ee

The PMNS matrix expresses the relationship between the neutrino
mass eigenstates and the flavor eigenstates, e.g.,
  \be
\left( \ba{c} \nu_e
\\ \nu_{\mu}
\\ \nu_{\tau} \ea \right)
  = \left( \ba{ccc} V_{e1} & V_{e2} & V_{e3} \\ V_{\mu 1} & V_{\mu 2} &
  V_{\mu 3} \\ V_{\tau 1} & V_{\tau 2} & V_{\tau 3} \ea \right)
 \left( \ba {c} \nu_1\\ \nu_2 \\ \nu_3 \ea \right),  \label{nm1}  \ee
where $ \nu_{e}$, $ \nu_{\mu}$, $\nu_{\tau}$ are the flavor
eigenstates, $ \nu_1$, $ \nu_2$, $ \nu_3$ are the mass eigenstates
and the $3 \times 3$ mixing matrix is the leptonic mixing matrix
\cite{pmns}. For the case of three Dirac neutrinos, in the
standard PDG parametrization \cite{pdg}, involving three angles
$\theta_{12}$, $\theta_{23}$, $\theta_{13}$ and the Dirac-like CP
violating phase ${\delta}_l$ the mixing matrix has the form
\begin{eqnarray}
V_{\rm PMNS}=   \left (
  \begin{array}{ccc}
    c_{12} c_{13} & s_{12} c_{13} & s_{13} e^{-i {\delta}_l} \\
    -s_{12} c_{23} - c_{12} s_{23} s_{13} e^{i {\delta}_l} & c_{12} c_{23} - s_{12}
    s_{23} s_{13}e^{i {\delta}_l} & s_{23} c_{13} \\
    s_{12} s_{23} - c_{12} c_{23} s_{13}e^{i {\delta}_l} & -c_{12} s_{23} - s_{12}
    c_{23} s_{13}e^{i {\delta}_l} & c_{23} c_{13}
  \end{array}
  \right ),
\label{ch1pmns2}
\end{eqnarray}
with $s_{ij} = {\rm sin}\theta_{ij}$, $c_{ij} = {\rm
cos}\theta_{ij}$. In the case of the Majorana neutrinos, there are
extra phases which cannot be removed. Therefore, the above matrix
takes the following form \beqn {\left( \ba{ccc} c_{12} c_{13} &
s_{12} c_{13} & s_{13}e^{-i {\delta}_l} \\ - s_{12} c_{23} -
c_{12} s_{23} s_{13} e^{i {\delta}_l} & c_{12} c_{23} - s_{12}
s_{23} s_{13} e^{i {\delta}_l} & s_{23} c_{13}
\\ s_{12} s_{23} - c_{12} c_{23} s_{13} e^{i \delta_l} & - c_{12}
s_{23} - s_{12} c_{23} s_{13} e^{i { \delta}_l} & c_{23} c_{13}
\ea \right)} \left( \ba{ccc} e^{i \alpha_1/2} & 0 & 0 \\ 0 &e^{i
\alpha_2/2} & 0 \\ 0 & 0  & 1 \ea \right), \label{nmm}\nonumber \\
\eeqn where $\alpha_1$ and $\alpha_2$ are the Majorana phases
which do not play any role in neutrino oscillations.

\section{Experimental status of fermion masses and mixing parameters
\label{inputs}}

To carry out any analysis regarding exploring the compatibility of
fermion mass matrices with the recent data, one needs to keep in
mind the experimental constraints imposed by the relationship
between mass matrices and their corresponding mixing matrices. To
facilitate our discussion in this regard, we first present the
status of relevant data in the quark as well as in the lepton
sector.

\subsection{Quark sector}
In the quark sector, the most important constraints are provided
by the directly observed quantities, such as masses of the quarks,
\vckm elements, CP violating phase ${\delta}$ and ${\rm
sin}2\beta$, etc.. The quark masses relevant for the present work
are the ``current'' quark masses at $m_Z$ energy scale
\cite{xingmass}, e.g., \beqn m_u = 1.27^{+0.5}_{-0.42}\, {\rm
MeV},~~~~~m_d = 2.90^{+1.24}_{-1.19}\, {\rm MeV},~~~~
m_s=55^{+16}_{-15}\, {\rm MeV},~~~~~~~~\nonumber\\ ~~~~~m_c=0.619
\pm 0.084\, {\rm GeV},~~ m_b=2.89 \pm 0.09\, {\rm GeV},~~
m_t=171.7 \pm 3.0\, {\rm GeV}. ~~~\label{qmasses} \eeqn The light
quark masses $m_u$, $m_d$ and $m_s$ are usually further
constrained by using the following mass ratios \cite{leut}
 \beqn ~~~~~~m_u / m_d =0.553 \pm
0.043~ , \,~~~~~~~~~~~~~m_s / m_d =18.9 \pm
 0.8 \;.
 \label{ratios} \eeqn

The analysis of mass matrices yields the CKM mixing matrix
elements which can then be compared with mixing matrix obtained
from a rigorous data based analysis. For ready reference as well
as for the sake of readability, we reproduce here the CKM matrix
as per PDG 2010 \cite{pdg}, at 95$\%$ C.L. as
  \be
   V_{\rm CKM} = \left( \ba {ccc} 0.97428\pm0.00015 &
   0.2253\pm0.0007 & 0.00347_{-0.00012}^{+0.00016} \\ 0.2252\pm0.0007 & 0.97345_{-0.00016}
^{+0.00015} & 0.0410_{-0.0007}^{+0.0011} \\
    0.00862_{-0.00020}^{+0.00026}& 0.0403_{-0.0007}^{+0.0011} &
0.999152_{-0.000045}^{+0.000030}\ea
    \right)\label{vckm}.
    \ee
Similarly, the precisely measured values \cite{pdg} of the CP
violating parameter sin$\,2\beta$, the Jarlskog rephasing
invariant parameter ${J}$ and the CP violating phase ${\delta}$
respectively are
\be
{\rm sin}2\beta=0.673 \pm 0.023,~~
J=(2.91_{-0.11}^{+0.19})\times10^{-5},~~
\delta=(73_{-25}^{+22})^\circ. \label{sin}\ee

\subsection{Leptonic sector\label{lepinputs}}
Adopting the three neutrino framework, several authors
\cite{kamland}-\cite{fogli2649} have presented updated information
regarding neutrino mass and mixing parameters obtained by carrying
out detailed global analyses. The latest situation regarding these
parameters at 3$\sigma$ C.L. is summarized as follows
\cite{schwetztortolavalle},
\be
 \Delta m_{21}^{2} = (7.05 - 8.34)\times
 10^{-5}~\rm{eV}^{2},~~~~
 |\Delta m_{31}^{2}| = (2.07 - 2.75)\times 10^{-3}~ \rm{eV}^{2},
 \label{solatmmass}\ee
\be
{\rm sin}^2\,\theta_{12}  =  0.25 - 0.37,~~~
 {\rm sin}^2\,\theta_{23}  =  0.36 - 0.67,~~~
 {\rm sin}^2\,\theta_{13} \leq 0.056. \label{s13}
\ee For the sake of completion as well as for ready reference, we
present the following PMNS matrix determined by taking into
account the neutrino oscillation data by Garcia {\it et al.}
\cite{garcia} at 3$\sigma$ C.L. as \be {\rm V_{PMNS}} =
\left(\ba{ccc}
  0.79~-~0.86~~ & 0.50~ -~ 0.61~~ & 0.00~-~0.20\\
 0.25 ~-~ 0.53~~  & 0.47 ~-~ 0.73~~  & 0.56~ -~0.79\\
 0.21 ~-~ 0.51~~ & 0.42 ~-~ 0.69~~ & 0.61~-~0.83
 \ea \right). \label{gar2} \ee

\section{Texture specific quark mass matrices \label{quarkmm}}

\subsection{Texture 6 zero matrices}

\begin{table}
{\bt{|c|c|c|} \hline
  & Class I  & Class II \\ \hline
a & $\left ( \ba{ccc} {\bf 0} & Ae^{i\alpha} & {\bf 0} \\
Ae^{-i\alpha}  & {\bf 0} & Be^{i\beta} \\ {\bf 0} & Be^{-i\beta} &
C \ea \right )$  & $\left ( \ba{ccc} {\bf 0} & Ae^{i\alpha} & {\bf
0} \\ Ae^{-i\alpha}  & D & {\bf 0} \\ {\bf 0} & {\bf 0}  & C \ea
\right )$ \\ b &  $\left ( \ba{ccc} {\bf 0} &{\bf 0} &
Ae^{i\alpha} \\ {\bf 0}  & C & Be^{i\beta} \\Ae^{-i\alpha} &
B^{-i\beta}  & {\bf 0} \ea \right )$  &
 $\left ( \ba{ccc} {\bf 0} & {\bf 0} & Ae^{i\alpha}
 \\ {\bf 0}  & C & {\bf 0} \\ Ae^{-i\alpha}  & {\bf 0}  &
D \ea \right )$ \\ c &  $\left ( \ba{ccc} {\bf 0} & Ae^{i\alpha} &
Be^{i\beta} \\ Ae^{-i\alpha}  & {\bf 0} & {\bf 0} \\ Be^{-i\beta}
& {\bf 0}  & C \ea \right )$  &
 $\left ( \ba{ccc} D & Ae^{i\alpha} &
{\bf 0} \\ Ae^{-i\alpha}  & {\bf 0} & {\bf 0} \\ {\bf 0} & {\bf 0}
& C \ea \right )$ \\ d &  $\left ( \ba{ccc} C & Be^{i\beta} & {\bf
0}
\\ Be^{-i\beta}  & {\bf 0} & Ae^{i\alpha}\\ {\bf 0}  & Ae^{-i\alpha} &
{\bf 0} \ea \right )$  &
 $\left ( \ba{ccc} C & {\bf 0} & {\bf 0}
 \\ {\bf 0}  & D & Ae^{i\alpha} \\ {\bf 0} & Ae^{-i\alpha}  &
{\bf 0} \ea \right )$ \\ e &  $\left ( \ba{ccc} {\bf 0} &
Be^{i\beta}  & Ae^{i\alpha} \\ Be^{-i\beta}  & C & {\bf 0} \\
Ae^{-i\alpha}  & {\bf 0}  &
 {\bf 0} \ea \right )$  &
 $\left ( \ba{ccc} D & {\bf 0} & Ae^{i\alpha}
\\ {\bf 0} & C &  {\bf 0} \\ Ae^{-i\alpha}  & {\bf 0}  &
{\bf 0}  \ea \right )$ \\ f & $\left ( \ba{ccc} C & {\bf 0} &
Be^{i\beta}
 \\ {\bf 0}  & {\bf 0}  & Ae^{i\alpha} \\Be^{-i\beta} & Ae^{-i\alpha}  &
{\bf 0} \ea \right )$  &
 $\left ( \ba{ccc} C & {\bf 0} &{\bf 0}
 \\ {\bf 0}  & {\bf 0} & Ae^{i\alpha} \\ {\bf 0} & Ae^{-i\alpha}  &
D \ea \right )$ \\    \hline \et}
 \caption{Twelve possibilities of
texture 3 zero mass matrices categorized into two classes I and
II, with each class having six matrices.} \label{t1}
\end{table}

As mentioned earlier, texture 6 zero Fritzsch mass matrices have
already been ruled out \cite{frzans,rrr}. Therefore, for the sake
of completion, we would like to discuss all possible non
Fritzsch-like combinations of texture 6 zero Hermitian mass
matrices as well. Before counting all possibilities, in view of
non zero masses of quarks, these matrices have to satisfy the
conditions ${\rm Trace}~ M_{U,D} \neq 0$ and ${\rm Det}~ M_{U,D}
\neq 0$. One can easily check that in case of texture 3 zero mass
matrices we arrive at 20 different possible texture patterns, out
of which 8 are easily ruled out by imposing these conditions. The
remaining 12 possible textures break into two classes, as shown in
Table (\ref{t1}), depending upon the equations these matrices
satisfy. For example, six matrices of class I, mentioned in Table
(\ref{t1}), satisfy the following equations
\be
C = m_1- m_2+ m_3, \quad A^2 + B^2 = m_1m_2 + m_2m_3 -m_1m_3,
\quad A^2 C = m_1m_2m_3. \label{classI}\ee
 Similarly, in case of class II,
all six matrices satisfy the following equations
\be
C + D= m_1 -m_2 +m_3,\quad A^2 -C D = m_1m_2 + m_2m_3 - m_1m_3,
\quad A^2 C = m_1m_2m_3. \label{classII}\ee
 The subscripts U and D have not been
used as these are valid for both kind of mass matrices. It may be
added that these classes are also related through permutation
symmetry \cite{brancoperm}.

Matrices $M_{U}$ and $M_{D}$ each can correspond to any of the 12
possibilities, therefore yielding 144 possible combinations which
in principle can yield 144 quark mixing matrices. These 144
combinations can be put into 4 different categories, e.g., if
$M_{U}$ is any of the 6 matrices from class I, then $M_{D}$ can be
either from class I or class II yielding 2 categories of 36
matrices each. Similarly, we obtain 2 more categories of 36
matrices each when $M_{U}$ is from class II and $M_{D}$ is either
from class I or class II. The 36 combinations in each category
further can be shown to be reduced to groups of six combinations
of mass matrices, each yielding same CKM matrix. Thus, the problem
of exploring the compatibility of 144 phenomenologically allowed
texture 6 zero combinations with the recent low energy data is
reduced only to an examination of 4 groups each having 6
combinations of mass matrices corresponding to the same CKM
matrix. Detailed analysis \cite{neeluijmpa} of these four groups
of texture 6 zero mass matrices reveals that all possible
combinations of texture 6 zero are ruled out as these are not able
to reproduce the CKM element $|V_{cb}|$.

\subsection{Texture 5 zero matrices}

Considering now the case of texture 5 zero mass matrices which
consist either of $M_{U}$ being 2 zero and $M_{D}$ being 3 zero or
vice versa. Texture 3 zero possibilities have already been
enumerated, therefore we consider only the possible patterns of
texture 2 zero mass matrices. After taking into consideration the
Trace and Determinant conditions mentioned earlier, one can check
that there are 18 possible texture 2 zero patterns. These textures
further break into three classes, detailed in Table (2), depending
upon the diagonalization equations satisfied by these matrices,
however, it can be shown that the classes IV and V  essentially
reduce to texture 3 zero patterns. We are therefore left with only
class III of texture 2 zero matrices that needs to be explored for
texture 5 zero combinations. All matrices of this class satisfy
the following equation
\be
 C + D = m_1- m_2+ m_3,\quad A^2 + B^2 - C D = m_1m_2 + m_2m_3 -m_1m_3,\quad A^2 C = m_1m_2m_3.
\label{classIII}\ee

Considering class III of texture 2 zero mass matrices along with
different patterns of class I and class II of texture 3 zero mass
matrices we find a total of 144 possibilities of texture 5 zero
mass matrices, in sharp contrast to the case if we had considered
the classes IV and V also yielding 432 possibilities. Keeping in
mind the hierarchy of the elements of the CKM matrix, we observe
that out of 144 cases, we are again left with only 4 such groups
of texture 5 zero mass matrices leading to mixing matrix having
hierarchical structure as that of CKM matrix.

\begin{table}
{\bt{|c|c|c|c|} \hline
 & Class III  & Class IV  & Class V   \\  \hline
a & $\left ( \ba{ccc} {\bf 0} & Ae^{i\alpha} & {\bf 0} \\
Ae^{-i\alpha}  & D & Be^{i\beta} \\ {\bf 0} & Be^{-i\beta}  & C
\ea \right )$  &
 $\left ( \ba{ccc} D & Ae^{i\alpha} &
{\bf 0} \\ Ae^{-i\alpha}  & {\bf 0} &  Be^{i\beta} \\ {\bf 0} &
Be^{-i\beta}  & C \ea \right )$     &
 $\left ( \ba{ccc} {\bf 0} & Ae^{i\alpha} &
 Fe^{i\gamma}\\ Ae^{-i\alpha}  &{\bf 0}  & Be^{i\beta} \\ Fe^{-i\gamma} & Be^{-i\beta}  &
C \ea \right )$ \\ b &  $\left ( \ba{ccc} {\bf 0} & {\bf 0}  &
Ae^{i\alpha} \\ {\bf 0}  & C & Be^{i\beta} \\  Ae^{-i\alpha} &
Be^{-i\beta}  & D \ea \right )$  &
 $\left ( \ba{ccc} D & {\bf 0} & Ae^{i\alpha}
 \\ {\bf 0} & C & Be^{i\beta} \\  Ae^{-i\alpha} & Be^{-i\beta}  &
{\bf 0} \ea \right )$     & $\left ( \ba{ccc} {\bf 0} &
Fe^{i\gamma} & Ae^{i\alpha}
\\  Fe^{-i\gamma}  & C  & Be^{i\beta} \\  Ae^{-i\alpha} & Be^{-i\beta}  &
 {\bf 0}\ea \right )$ \\
c &  $\left ( \ba{ccc} D & Ae^{i\alpha} &
  Be^{i\beta}\\ Ae^{-i\alpha}  & {\bf 0}  & {\bf 0} \\  Be^{-i\beta} & {\bf 0}
& C \ea \right )$  & $\left ( \ba{ccc} {\bf 0} & Ae^{i\alpha} &
 Be^{i\beta} \\ Ae^{-i\alpha}  & D & {\bf 0}  \\  Be^{-i\beta}  & {\bf 0} &
C \ea \right )$     & $\left ( \ba{ccc} {\bf 0} & Ae^{i\alpha} &
Be^{i\beta}
\\ Ae^{-i\alpha} & {\bf 0} & Fe^{i\gamma} \\ Be^{-i\beta}  & Fe^{-i\gamma} &
C \ea \right )$ \\ d &  $\left ( \ba{ccc} C & Be^{i\beta} & {\bf
0}
 \\ Be^{-i\beta} & D & Ae^{i\alpha}  \\ {\bf 0} & Ae^{-i\alpha} & {\bf 0}
 \ea \right )$  &
 $\left ( \ba{ccc} C & Be^{i\beta} &  {\bf 0}
 \\ Be^{-i\beta} & {\bf 0} & Ae^{i\alpha} \\ {\bf 0} & Ae^{-i\alpha}  &
D \ea \right )$     &
 $\left ( \ba{ccc} C & Be^{i\beta}  & Fe^{i\gamma}
\\ Be^{-i\beta}   & {\bf 0}  &  Ae^{i\alpha} \\ Fe^{-i\gamma} & Ae^{-i\alpha}  &
 {\bf 0}\ea \right )$ \\
e &  $\left ( \ba{ccc} D & Be^{i\beta} & Ae^{i\alpha}
  \\ Be^{-i\beta}  & C & {\bf 0} \\ Ae^{-i\alpha} & {\bf 0}
& {\bf 0} \ea \right )$  &
 $\left ( \ba{ccc} {\bf 0} & Be^{i\beta} & Ae^{i\alpha}
  \\  Be^{-i\beta}  & C & {\bf 0}  \\ Ae^{-i\alpha}   & {\bf 0} &
D \ea \right )$     & $\left ( \ba{ccc} {\bf 0}  & Be^{i\beta} &
Ae^{i\alpha}
\\ Be^{-i\beta}  & C & Fe^{i\gamma} \\ Ae^{-i\alpha} & Fe^{-i\gamma}&
{\bf 0}  \ea \right )$ \\ f &  $\left ( \ba{ccc} C & {\bf 0} &
Be^{i\beta}
 \\ {\bf 0} & {\bf 0} & Ae^{i\alpha}  \\ Be^{-i\beta} & Ae^{-i\alpha} & D
 \ea \right )$  &
$\left ( \ba{ccc} C  &  {\bf 0} & Be^{i\beta}
 \\ {\bf 0} & D & Ae^{i\alpha} \\Be^{-i\beta}  & Ae^{-i\alpha}  &
0 \ea \right )$     &
 $\left ( \ba{ccc} C & Fe^{i\gamma} & Be^{i\beta}
\\  Fe^{-i\gamma} & {\bf 0}  &  Ae^{i\alpha} \\ Be^{i\beta} & Ae^{-i\alpha}  &
 {\bf 0}\ea \right )$ \\  \hline
\et} \caption{Texture 2 zero possibilities categorized into three
classes III, IV and V, with each class having six matrices.}
 \label{t2}
\end{table}

A detailed analysis of these texture 5 zero mass matrices have
been carried out \cite{neeluijmpa} which shows that interestingly
only one possibility, corresponding to the usual Fritzsch-like
texture 5 zero mass matrix where $M_{U}$ is of texture 2 zero and
$M_{D}$ is of texture 3 zero type, appears to be viable. Also it
may be added that the viability of this combination depends on the
light quark masses used as inputs.

\subsection{Texture 4 zero matrices}
As is well known, Fritzsch-like texture 4 zero mass matrices are
compatible with the quark mixing data, however, there are two
issues which need to be addressed in this context. Firstly, one
needs to carry out a detailed analysis of the various possible non
Fritzsch-like texture 4 zero mass matrices. In fact, the texture
two zero possibilities, presented in Table (2), result into 324
texture 4 zero possibilities, the analysis of such a large number
of possibilities is yet to be carried out. The other issue is
whether we can consider `weakly' hierarchical mass matrices to
reproduce `strongly' hierarchical mixing angles. This issue has
been explored in a detailed manner \cite{cps,s2b} and in the
present work we reproduce some of the essential details regarding
this.

In this context, on the one hand, many authors
\cite{hallraisin}-\cite{branco} have shown that on using strong
hierarchy of the elements of the mass matrices, texture 4 zero
mass matrices appear to be incompatible with the recent value of
sin$\,2\beta$. While on the other hand, recently \cite{cps,s2b}
extension of the parameter space of the elements of these matrices
has been carried out to include the case of `weak hierarchy'
amongst them along with the usually considered `strong hierarchy'
case and thereby they have been shown to be compatible with the
parameter sin$\,2\beta$.

It may be noted that although hierarchy of the elements of the
mass matrices has been considered often while carrying out the
analysis, however it has been explicitly defined only in
\cite{s2b}. The various relations between the elements of the mass
matrices, given in Eq. (\ref{nf2zero}), $A_i , B_i, C_i, D_i$ $(i
= U, D)$ essentially correspond to the structural features of the
mass matrices including their hierarchies. As is usual the element
$|A_i|$ takes a value much smaller than the other three elements
of the mass matrix which can assume different relations amongst
each other, defining different hierarchies. For example, in case
$D_i < |B_{i}| < C_i$ it would lead to a strongly hierarchical
mass matrix whereas a weaker hierarchy of the mass matrix implies
$D_{i} \lesssim |B_i| \lesssim C_i$. It may also be added that for
the purpose of numerical work, one can conveniently take the ratio
$D_i/C_i \sim 0.01$ characterizing strong hierarchy whereas
$D_i/C_i \gtrsim 0.2$ implying weak hierarchy.

The analysis carried out by \cite{cps} incorporates the quark
masses and their ratios mentioned in Eqs. (\ref{qmasses})and
(\ref{ratios}) as well as by imposing the constraints given in
Eqs. (\ref{vckm}) and (\ref{sin}). Further, full variation has
been given to the phases associated with the mass matrices
$\phi_1$ and $\phi_2$, the parameters $D_U$ and $D_D$ have been
given wide variation in conformity with the hierarchy of the
elements of the mass matrices e.g., $D_i < C_i$ for $i=U, D$. The
extended range of these parameters allows the calculations for the
case of weak hierarchy of the elements of the mass matrices as
well.

To begin with, in Fig. 1(a) $C_{U }/m_t$ versus $C_{ D}/m_b$ has
been plotted. A look at the figure reveals that both $C_{U}/m_t$
as well as $C_{ D}/m_b$ take values from $\sim 0.55-0.95$, which
interestingly indicates the ratios being almost proportional.
Also, the figure gives interesting clues regarding the role of
strong and weak hierarchy. In particular, one finds that in case
one restricts to the assumption of strong hierarchy then these
ratios take large values around $0.95$. However, for the case of
weak hierarchy, the ratios $C_{U }/m_t$ and $C_{ D}/m_b$ take much
larger number of values, in fact almost the entire range mentioned
above, which are compatible with the data.

\begin{figure}[hbt]
\vspace{0.12in}
\centerline{\epsfysize=1.55in\epsffile{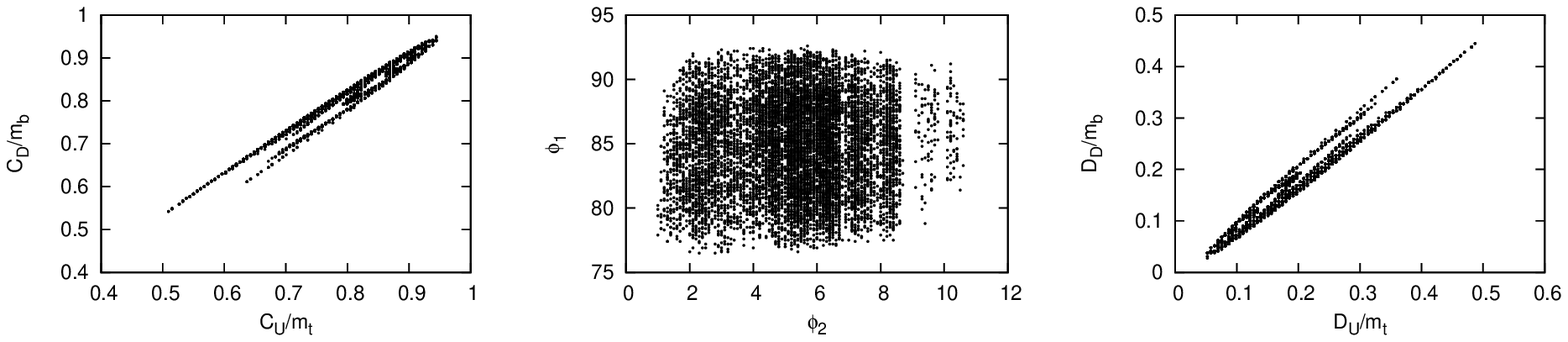}}
\vspace{0.08in}
   \caption{Plots showing the allowed ranges of (a) $C_{\rm U}/m_t$ versus $C_{\rm
D}/m_b$, (b) $\phi_1$ versus $\phi_2$ and (c) $D_{\rm U}/m_t$
versus $D_{\rm D}/m_b$}
  \label{cpsfigs}
  \end{figure}

In Fig. 1(b), the plot of $\phi_1$ versus $\phi_2$ has been
presented. Interestingly, the present refined inputs limit the
ranges of the two phases to $\phi_1 \sim 76^{\rm o} - 92^{\rm o}$
and $\phi_2 \sim 1^{\rm o} - 11^{\rm o}$. Keeping in mind that
full variation has been given to the free parameters $D_U$ and
$D_D$, corresponding to both strong as well as weak hierarchy
cases, it may be noted that the allowed ranges of the two phases
come out to be rather narrow. In particular, for the strong
hierarchy case one gets $\phi_2 \sim 10^{\rm o}$, whereas for the
case of weak hierarchy $\phi_2$ takes almost its entire range
mentioned above. Also, the analysis indicates that although
$\phi_1 \gg \phi_2$, still both the phases are required for
fitting the mixing data.

As a next step, the role of the hierarchy defining parameters
$D_{U}$ and $D_{D}$ has been emphasized. To this end, in Fig. 1(c)
$D_{U }/m_t$ versus $D_{ D}/m_b$ has been given, representing an
extended range of the parameters $D_U$ and $D_D$. A closer look at
the figure reveals both $D_{U}/m_t$ as well as $D_{ D}/m_b$ take
values $\sim 0.05-0.5$. The lower limit of the range i.e. when the
ratios $D_{U}/m_t$ and $D_{ D}/m_b$ are around $0.05$ corresponds
to strong hierarchy amongst the elements of the mass matrices,
whereas when the elements have weak hierarchy then these ratios
take a much larger range of values. From this one may conclude
that in the case of strongly hierarchical elements of the texture
4 zero mass matrices, one has limited compatibility of these
matrices with the quark mixing data, whereas the weakly
hierarchical ones indicate the compatibility for much broader
range of the elements.

\section{Texture specific lepton mass matrices \label{lepmm}}

\subsection{Texture 6 zero matrices}
Having discussed the texture specific quark mass matrices, in the
light of quark lepton unification hypothesis advocated by Smirnov
\cite{qlepuni}, one would also like to know the status of texture
6 zero Fritzsch as well as non Fritzsch-like lepton mass matrices.
A detailed analysis of texture 6 zero mass matrices have been
carried out by several authors \cite{ourneut6zero,neelu6zero}. In
particular, for normal hierarchy of neutrino masses Zhou and Xing
\cite{zhou} have carried out an analysis of all possible Fritzsch
as well as non Fritzsch-like texture 6 zero lepton mass matrices.
Their analysis has been extended further to include inverted
hierarchy and non degenerate scenario of neutrino masses
\cite{neelu6zero}.

As already shown for the case of quarks in Section(\ref{quarkmm}),
there are a total number of 144 possible cases of texture 6 zero
mass matrices. For the case of lepton mass matrices, there are 6
cases for each of the 144 combinations corresponding to normal/
inverted hierarchy and degenerate scenario of neutrino masses for
Majorana neutrinos as well as Dirac neutrinos, leading to a total
of 864 cases. The analysis carried out in \cite{neelu6zero}
reveals several interesting points. In particular, their
investigations for Dirac neutrinos show that there are no viable
texture 6 zero lepton mass matrices for normal/ inverted hierarchy
as well as degenerate scenario of neutrino masses. For the case of
Majorana neutrinos for texture 6 zero lepton mass matrices, again
all the cases pertaining to inverted hierarchy and degenerate
scenario of neutrino masses are also ruled out. Assuming normal
hierarchy of Majorana neutrinos, the analysis reveals that out of
144, only 16 combinations are compatible with current neutrino
oscillation data at $3\sigma$ C.L..

\subsection{Texture 5 zero matrices}

Similar to the case of texture 6 zero lepton mass matrices, the
implications for different hierarchies in the case of texture 5
zero lepton mass matrices have also been investigated for both
Majorana and Dirac neutrinos \cite{neeluunpublished}. For the two
types of neutrinos, corresponding to normal/ inverted hierarchy
and degenerate scenario of neutrino masses 360 cases each have
been considered for carrying out the analysis, making it a total
of 2160 cases.

For Majorana neutrinos with normal hierarchy of neutrino masses,
out of the 360 combinations, 67 are compatible with the neutrino
mixing data. Most of the phenomenological implications of
combinations of different categories are similar, however, still
these can be experimentally distinguished with more precise
measurements of $\theta_{13}$ and $\theta_{23}$. Interestingly,
degenerate scenario of Majorana neutrinos is completely ruled out
by the existing data. In the case of inverted hierarchy, 24
combinations out of 360 are compatible with the neutrino mixing
data.

For Dirac neutrinos with normal hierarchy of neutrino masses, as
compared to Majorana cases, out of 360 only 44 combinations are
compatible with neutrino mixing data. Interestingly, 6
combinations out of 44 can accommodate degenerate Dirac neutrinos.
For inverted hierarchy, 24 combinations are compatible with the
existing data.

\subsection{Texture 4 zero matrices}
Like the case of quarks, the number of viable possibilities for
the case of texture 4 zero lepton mass matrices is also quite
large, so in the present work we have discussed the essentials of
recent detailed analyses \cite{ourneut4zero} regarding only the
Fritzsch-like texture 4 zero mass matrices. In particular, they
have investigated the implications of different hierarchies of
neutrino masses on these matrices for both Majorana and Dirac
neutrinos. Interestingly, at 3$\sigma$ C.L., their analysis rules
out both inverted hierarchy and degenerate scenario of neutrino
masses for the two types of neutrinos. We reproduce here the
essentials of their arguments.

\begin{figure}[hbt]
\vspace{0.12in}
\centerline{\epsfysize=2.8in\epsffile{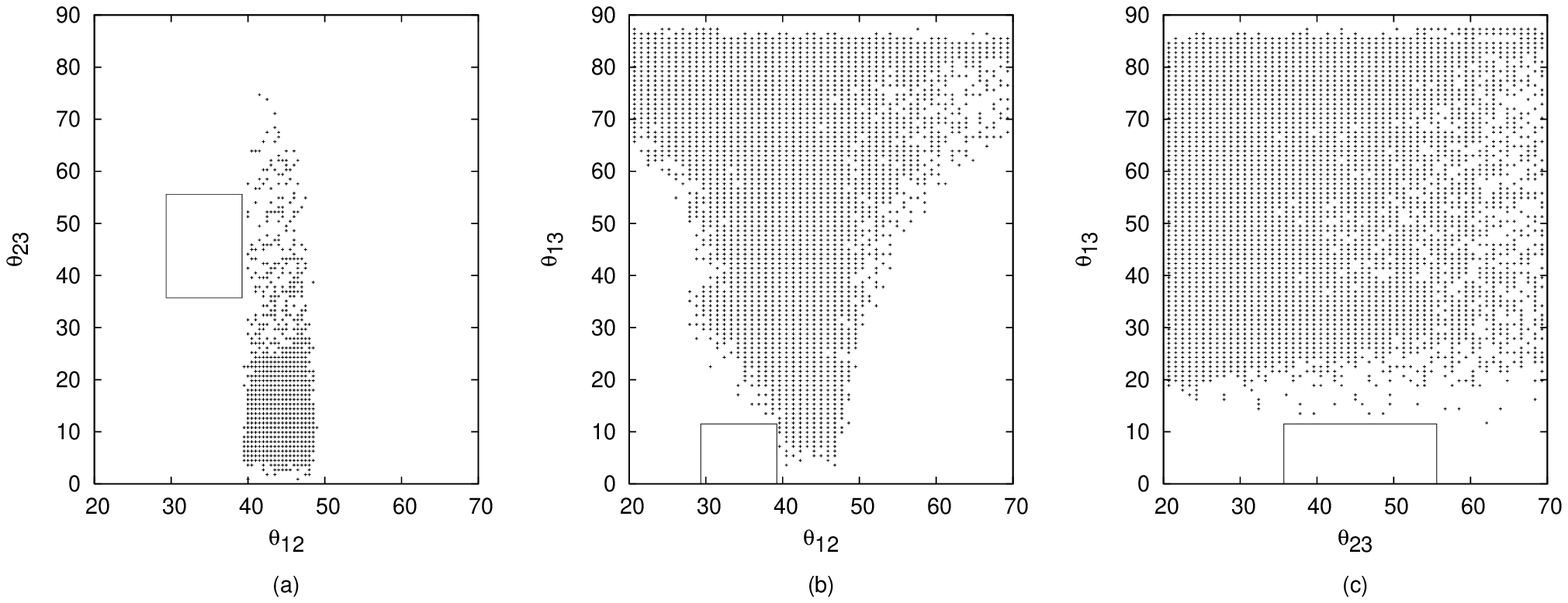}}
\vspace{0.08in}
   \caption{Plots showing the parameter space
corresponding to any of the two mixing angles by constraining the
third angle by its experimental limits and giving full allowed
variation to other parameters for Majorana neutrinos. The blank
rectangular region indicates the experimentally allowed $3\sigma$
region of the plotted angles.}
  \label{fig1}
  \end{figure}

Basically, they have plotted the parameter space corresponding to
any of the two mixing angles by constraining the third angle by
its experimental values, mentioned in Eq. (\ref{s13}), while
giving full allowed variation to other parameters. For ready
reference we present these graphs in  Fig. (\ref{fig1}). These
plots immediately reveal that the inverted hierarchy is ruled out
at 3$\sigma$ C.L.. For example, from Fig. (\ref{fig1}a),
(\ref{fig1}b) and (\ref{fig1}c), for Majorana neutrinos, one can
note that the plotted parameter space of the angles has no overlap
with their experimentally allowed $3\sigma$ region. Similar plots
pertaining to Dirac neutrinos also rule out inverted hierarchy of
neutrino masses.

Their analysis also shows that for Majorana or Dirac neutrinos the
cases of neutrino masses being degenerate, characterized by either
$m_{\nu_1} \lesssim m_{\nu_2} \sim m_{\nu_3} \lesssim 0.1~\rm{eV}$
or $m_{\nu_3} \sim m_{\nu_1} \lesssim m_{\nu_2} \lesssim
0.1~\rm{eV}$ corresponding to normal and inverted hierarchy
respectively, are again ruled out. Considering degenerate scenario
corresponding to inverted hierarchy, Fig. (\ref{fig1}) can again
be used to rule out degenerate scenario at 3$\sigma$ C.L. for
Majorana neutrinos. It needs to be mentioned that while plotting
these figures the range of the lightest neutrino mass is taken to
be $10^{-8}\,\rm{eV}-10^{-1}\,\rm{eV}$, which includes the
neutrino masses corresponding to degenerate scenario, therefore by
discussion similar to the one given for ruling out inverted
hierarchy, degenerate scenario of neutrino masses is ruled out as
well.

\begin{figure}[tbp]
\centerline{\epsfysize=2.8in\epsffile{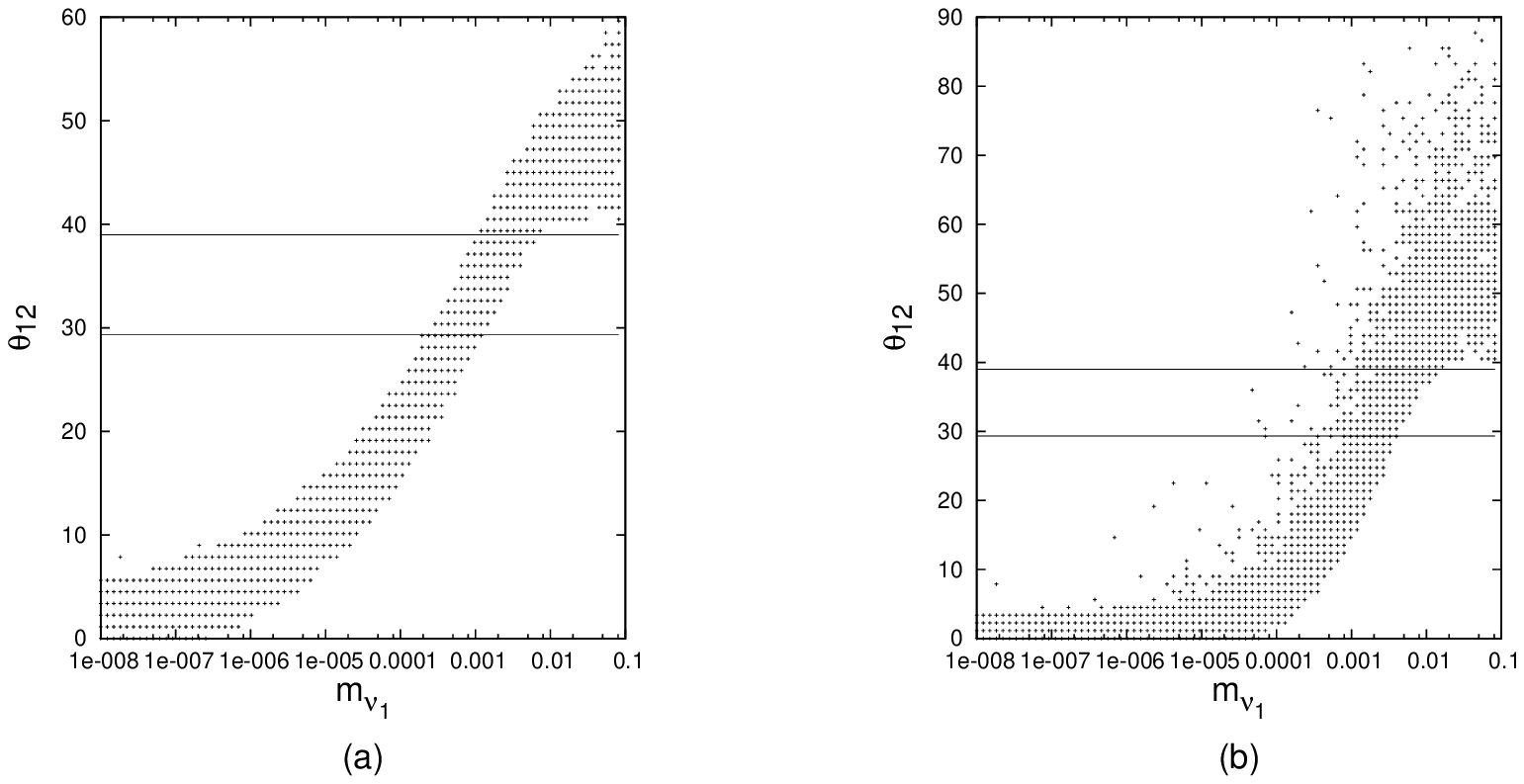}}
\vspace{0.08in}
 \caption{Plots showing variation of mixing angle
$\theta_{12}$ with lightest neutrino mass $m_{\nu_1}$ by giving
full variation to other parameters for (a) Majorana neutrinos and
(b) Dirac neutrinos. The parallel lines indicate the $3\sigma$
limits of angle $\theta_{12}$.}
  \label{th12vm-md}
  \end{figure}

Coming to degenerate scenario corresponding to normal hierarchy,
one can easily show that this is ruled out again. To this end, in
Fig. (\ref{th12vm-md}), by giving full variation to other
parameters, the plot of the mixing angle $\theta_{12}$ against the
lightest neutrino mass $m_{\nu_1}$ have been presented. Fig.
(\ref{th12vm-md}a) corresponds to the case of Majorana neutrinos
and Fig. (\ref{th12vm-md}b) to the case of Dirac neutrinos. From
the figures one can immediately find that the values of
$\theta_{12}$ corresponding to $m_{\nu_1} \lesssim 0.1~\rm{eV}$
lie outside the experimentally allowed range, thereby ruling out
degenerate scenario for Majorana as well as Dirac neutrinos at
3$\sigma$ C.L..

After ruling out the cases pertaining to inverted hierarchy and
degenerate scenarios, we come the normal hierarchy cases for
Majorana as well as Dirac neutrinos. For both types of neutrinos,
these yield viable ranges of neutrino masses, mixing angles
$\theta_{12}$, $\theta_{23}$ and $\theta_{13}$, Jarlskog's
rephasing invariant parameter in the leptonic sector $J_l$ and the
Dirac-like CP violating phase in the leptonic sector $\delta_l$
The analysis reveals several interesting points. For both Dirac or
Majorana neutrinos, the viable range of the lightest neutrino mass
$m_{\nu_1}$ is quite different, in particular the range
corresponding to Dirac neutrinos is much wider at both the ends as
compared to the Majorana neutrinos. Therefore, a measurement of
$m_{\nu_1}$ could have important implications for the nature of
neutrinos. Also, one finds that the lower limit on $\theta_{13}$
for the Dirac case is considerably lower than for the Majorana
case, therefore a measurement of $\theta_{13}$ would have
important implications for this case. The different cases of Dirac
and Majorana neutrinos do not show any divergence for the ranges
of Jarlskog's rephasing invariant parameter.

\section{Summary and Conclusion \label{summ}}
Fritzsch-like texture specific mass matrices have provided
important clues for understanding the pattern of quark mixings and
CP violation. Likewise, in the leptonic sector also texture
specific mass matrices are useful in explaining the pattern of
neutrino masses and mixings. To tackle the larger issue of quark
and lepton mixing phenomena together, it is perhaps desirable to
take into account the quark-lepton unification hypothesis
\cite{qlepuni}. This immediately brings forth the issue of finding
the simplest texture structure at the leading order, compatible
with the quark and lepton mixing phenomena. Further, in the
absence of any theoretical justification for Fritzsch-like mass
matrices, one also needs to consider non Fritzsch-like mass
matrices for quarks as well as leptons.

In the present work, we have given an overview of all possible
cases of Fritzsch-like as well as non Fritzsch-like texture 6 and
5 zero fermion mass matrices, for details see
\cite{ourneut6zero,neeluijmpa}. Further, for the case of texture 4
zero Fritzsch-like quark mass matrices, the issue of the hierarchy
of the elements of the mass matrices and the role of their phases
have been discussed, details can be found in \cite{cps,s2b}.
Furthermore, the case of texture 4 zero Fritzsch-like lepton mass
matrices has also been discussed with an emphasis on the hierarchy
of neutrino masses for both Majorana and Dirac neutrinos,
elaborate analyses presented in \cite{ourneut4zero}.

These analyses reveal several interesting results. In principle,
for the case of quarks \cite{neeluijmpa}, there are 144
combinations of texture 6 zero mass matrices whereas in the case
of texture 5 zero matrices one can arrive at 360 combinations.
Interestingly, all the texture 6 zero combinations are completely
ruled out whereas in the case of texture 5 zero mass matrices the
only viable possibility looks to be that of Fritzsch-like matrices
which shows only limited viability, depending upon the light quark
masses used as input.

Further, for the case of texture 4 zero quark mass matrices
\cite{cps,s2b}, including the case of `weak hierarchy' along with
the usually considered `strong hierarchy' case, one finds that the
weakly hierarchical mass matrices are able to reproduce the
strongly hierarchical mixing angles. Also, both the phases having
their origin in the mass matrices have to be non zero to achieve
compatibility of these matrices with the quark mixing data.

The same number of combinations have been investigated for the
neutrino mixing data considering normal/ inverted hierarchy and
degenerate scenario of neutrino masses for Majorana as well as
Dirac neutrinos. Texture 6 zero in the case of leptons results
into 864 cases to be analyzed \cite{neelu6zero}. Interestingly,
all the possibilities pertaining to normal/ inverted hierarchy and
degenerate scenario of neutrino masses for Dirac neutrinos and
inverted hierarchy as well as degenerate scenarios in the case of
Majorana neutrinos are ruled out. Normal hierarchy of neutrino
masses for Majorana neutrinos results into 16 combinations out of
144 which are in accordance with the neutrino oscillation data.

Texture 5 zero entails considering 2160 cases to ascertain their
compatibility with the neutrino mixing data
\cite{neeluunpublished}. Interestingly, one finds that texture 5
zero lepton mass matrices can accommodate all
 hierarchies of neutrino masses. In the case of normal hierarchy,
67 combinations for Majorana neutrinos and 44 combinations for
Dirac neutrinos are compatible with the neutrino mixing data.
There are 6 combinations, out of 44, which can accommodate
degenerate Dirac neutrinos. Further, in case of inverted
hierarchy, 24 combinations are compatible both for Majorana as
well as Dirac neutrinos.

For the Fritzsch-like texture 4 zero neutrino mass matrices
\cite{ourneut4zero}, analysis pertaining to both Majorana and
Dirac neutrinos for different hierarchies of neutrino masses
reveals that for both types of neutrinos, all the cases pertaining
to inverted hierarchy and degenerate scenarios of neutrino masses
are ruled out at $3\sigma$ C.L. by the existing data. For the
normal hierarchy cases, one gets viable ranges of neutrino masses,
mixing angle $s_{13}$, Jarlskog's rephasing invariant parameter
$J_l$ and the CP violating Dirac-like phase $\delta_l$.
Interestingly, a measurement of $m_{\nu_1}$ and mixing angle
$\theta_{13}$ could have important implications for the nature of
neutrinos.

In conclusion, we would like to remark that on the one hand there
is a need to take the analysis of texture specific mass matrices
towards completion. For example, besides carrying out the analysis
of texture 4 zero non Fritzsch-like fermion mass matrices, one has
to consider texture 3 zero cases also, the latter corresponding to
general mass matrices after carrying out weak basis rotations. On
the other hand, one may also consider breaking the hermiticity
condition perturbatively as has been done recently
\cite{frixingzhou} and to go into its detailed implications.
Similarly, the issue of phases of mass matrices and their
relationship with the CP violating parameters also needs a careful
look and detailed investigations.

\section*{Acknowledgments}
The authors would like to thank the organizers of the
``International Conference on Flavor Physics in the LHC Era" held
at Singapore in November 2010 for giving an opportunity to present
some aspects of the present work. The authors would also like to
thank the participants for several interesting discussions. G.A.
would also like to acknowledge DST, Government of India for
financial support and the Chairman, Department of Physics for
providing facilities to work in the department.


\begin{thebibliography}{00}
\bibitem{cabibbo}N. Cabibbo, {\it Phys. Rev. Lett.} {\bf 10}, 531
(1963).

\bibitem{glashow}S. L. Glashow, J. Illiopoulos and L. Maiani, {\it
Phys. Rev. D} {\bf 2}, 1285 (1970).

\bibitem{kobayashi}M. Kobayashi and T. Maskawa, {\it Prog. Theor. Phys.} {\bf 49}, 652 (1973).

\bibitem{horizontal}C. D. Froggatt and H. B. Nielsen, {\it Nucl. Phys. B}
{\bf 147}, 277 (1979); Y. Nir and N. Seiberg, {\it Phys. Lett. B}
{\bf 309}, 337 (1993); M. Leurer, Y. Nir and N. Seiberg, {\it
Nucl. Phys. B} {\bf 420}, 468 (1994); L. E. Ibanez and G. G. Ross,
{\it Phys. Lett. B} {\bf 332}, 100 (1994); G. Altarelli and F.
Feruglio, arXiv:1002.0211.

\bibitem{extradim}A. J. Buras, C. Grojean, S. Pokorski and R.
Ziegler, arXiv:1105.3725, and references therein; Z. Guo and B.
Ma, {\it JHEP} {\bf 0909}, 091 (2009), and references therein.

\bibitem{frzans}H. Fritzsch, {\it Phys. Lett. B} {\bf 70}, 436 (1977);
{\it ibid.} {\it Phys. Lett. B} {\bf 73}, 317 (1978).

\bibitem{ansatze}B. Stech, {\it Phys. Lett. B} {\bf 130}, 189
(1983); M. Gronau, R. Johnson and J. Schechter, {\it Phys. Rev.
Lett.} {\bf 54}, 2176 (1985).

\bibitem{rrr}P. Ramond, R. G. Roberts and G. G. Ross,  {\it Nucl.
Phys. B} {\bf 406}, 19 (1993).

\bibitem{fri2000}H. Fritzsch and Z. Z. Xing,
{\it Prog. Part. Nucl. Phys.} {\bf45}, 1 (2000), and references
therein.

\bibitem{group5zero}P. S. Gill and M. Gupta, {\it J. Phys. G: Nuc. Part.
Phys.} {\bf 21}, 1 (1995).

\bibitem{5zero}A. Rasin, {\it Phys. Rev. D} {\bf 58}, 096012
(1998); H. D. Kim and G. H. Wu, hep-ph/0004036; B. R. Desai and A.
R. Vaucher, {\it Phys. Rev. D} {\bf 63}, 113001 (2001).

\bibitem{group4zero}N. G. Deshpande, M. Gupta and P. B. Pal, {\it
Phys. Rev. D} {\bf 45}, 953 (1992); P. S. Gill and M. Gupta, {\it
Pramana} {\bf 45}, 333 (1995); {\it ibid.} {\it J. Phys. G} {\bf
23}, 335 (1997); {\it ibid.} {\it Phys. Rev. D} {\bf 56}, 3143
(1997); M. Randhawa and M. Gupta, {\it Phys. Rev. D} {\bf 63},
097301 (2001).

\bibitem{xingzhang}Z. Z. Xing and H. Zhang, {\it J. Phys. G} {\bf 30}, 129 (2004).

\bibitem{cps}R. Verma, G. Ahuja, N. Mahajan, M. Randhawa
and M. Gupta, {\it J. Phys. G: Nucl. Part. Phys.} {\bf 37}, 075020
(2010), and references therein.

\bibitem{hallraisin}L. J. Hall and A. Rasin, {\it Phys. Lett. B} {\bf 315}, 164 (1993).

\bibitem{barbieri}R. Barbieri, L. J. Hall and A. Romanino, {\it Phys. Lett. B} {\bf 401}, 47
(1997).

\bibitem{roberts}R. G. Roberts, A. Romanino, G. G. Ross and
L. Velasco-Sevilla, {\it Nucl. Phys. B} {\bf 615}, 358 (2001).

\bibitem{frixingt4s2b}H. Fritzsch and Z. Z. Xing, {\it Phys. Lett. B} {\bf 555}, 63 (2003).

\bibitem{kimraby}H. D. Kim, S. Raby and L. Schradin, {\it Phys. Rev. D} {\bf 69}, 092002 (2004).

\bibitem{branco}G. C. Branco, M. N. Rebelo and J. I. Silva-Marcos, {\it Phys. Rev.
D} {\bf 76}, 033008 (2007).

\bibitem{bando}M. Bando, S. Kaneko, M. Obara and M. Tanimoto, {\it Prog. Theor.
Phys.} {\bf 116}, 1105 (2007).

\bibitem{s2b}R. Verma, G. Ahuja and M. Gupta, {\it Phys. Lett. B} {\bf 681},
330 (2009).

\bibitem{framp}P. H. Frampton, S. L. Glashow and D. Marfatia,
{\it Phys. Lett. B} {\bf 536}, 79 (2002).

\bibitem{0307359}Z. Z. Xing, {\it Int. J. Mod. Phys. A} {\bf 19}, 1 (2004), and
references therein.

\bibitem{fuku}M. Fukugita, M. Tanimoto and T. Yanagida, {\it Phys. Rev. D} {\bf 57},
 4429 (1998), and references therein.

\bibitem{zhou}S. Zhou and Z. Z. Xing, {\it Eur. Phys. J. C} {\bf 38}, 495 (2005).

\bibitem{ourneut6zero}M. Randhawa, G. Ahuja and M. Gupta, {\it Phys. Lett. B} {\bf 643},
175 (2006).

\bibitem{ourneut4zero}G. Ahuja, S. Kumar, M. Randhawa, M. Gupta and S.
Dev, {\it Phys. Rev. D} {\bf 76}, 013006 (2007);G. Ahuja, M.
Gupta, M. Randhawa and R. Verma, {\it Phys. Rev. D} {\bf 79},
093006 (2009).

\bibitem{leptex}K. Kang and S. K. Kang, {\it Phys. Rev. D} {\bf 56}, 1511 (1997);
P. S. Gill and M. Gupta, {\it Phys. Rev. D} {\bf 57}, 3971 (1998);
M. Randhawa, V. Bhatnagar, P. S. Gill and M. Gupta, {\it Phys.
Rev. D} {\bf 60}, 051301 (1999); H. Nishiura, K. Matsuda and T.
Fukuyama, {\it Phys. Rev. D} {\bf 60}, 013006 (1999); K. Matsuda,
T. Fukuyama and H. Nishiura, {\it Phys. Rev. D} {\bf 61}, 053001
(2000); K. Kang, S. K. Kang, C. S. Kim and S. M. Kim, {\it Mod.
Phys. Lett. A} {\bf 16}, 2169 (2001); C. Giunti and M. Tanimoto,
{\it Phys. Rev. D} {\bf 66}, 113006 (2002); M. Frigerio and A. Yu.
Smirnov, {\it Nucl. Phys. B} {\bf 640}, 233 (2002); P. F. Harrison
and W. G. Scott, {\it Phys. Lett. B} {\bf 547}, 219 (2002); E. Ma,
{\it Mod. Phys. Lett. A} {\bf 17}, 2361 (2002); {\it ibid.} {\it
Phys. Rev. D} {\bf 66}, 117301 (2002); Z. Z. Xing, {\it Phys.
Lett. B} {\bf 530}, 159 (2002); {\it ibid.} {\it Phys. Lett. B}
{\bf 539}, 85 (2002); A. Kageyama, S. Kaneko, N. Simoyama and M.
Tanimoto, {\it Phys. Lett. B} {\bf 538}, 96 (2002); W. L. Guo and
Z. Z. Xing, hep-ph/0211315; M. Randhawa, G. Ahuja and M. Gupta,
{\it Phys. Rev. D} {\bf 65}, 093016 (2002); M. Frigerio and A. Yu.
Smirnov, {\it Phys. Rev. D} {\bf 67}, 013007 (2003); K. S. Babu,
E. Ma and J. W. F. Valle, {\it Phys. Lett. B} {\bf 552}, 207
(2003); W. L. Guo and Z. Z. Xing, {\it Phys. Rev. D} {\bf 67},
053002 (2003); S. Kaneko and M. Tanimoto, {\it Phys. Lett. B} {\bf
551}, 127 (2003); B. R. Desai, D. P. Roy and A. R. Vaucher, {\it
Mod. Phys. Lett. A} {\bf 18}, 1355 (2003); K. Hasegawa, C. S. Lim
and K. Ogure, {\it Phys. Rev. D} {\bf 68}, 053006 (2003); M.
Honda, S. Kaneko and M. Tanimoto, {\it JHEP} {\bf 0309}, 028
(2003); G. Bhattacharyya, A. Raychaudhuri and A. Sil, {\it Phys.
Rev. D} {\bf 67}, 073004 (2003); P. F. Harrison and W. G. Scott,
{\it Phys. Lett. B} {\bf 594}, 324 (2004); M. Bando, S. Kaneko, M.
Obara and M. Tanimoto, {\it Phys. Lett. B} {\bf 580}, 229 (2004);
O. L. G. Peres and A. Yu. Smirnov, {\it Nucl. Phys. B} {\bf 680},
479 (2004); C. H. Albright, {\it Phys. Lett. B} {\bf 599}, 285
(2004); J. Ferrandis and S. Pakvasa, {\it Phys. Lett. B} {\bf
603}, 184 (2004); S. T. Petcov and W. Rodejohann, {\it Phys. Rev.
D} {\bf 71}, 073002 (2005); S. S. Masood, S. Nasri and J.
Schechter, {\it Phys. Rev. D} {\bf 71}, 093005 (2005); R.
Derm\'i\v{s}ek and S. Raby, {\it Phys. Lett. B} {\bf 622}, 327
(2005); F. Plentinger and W. Rodejohann, {\it Phys. Lett. B} {\bf
625}, 264 (2005); S. Dev and S. Kumar, {\it Mod. Phys. Lett. A}
{\bf 22}, 1401 (2007); S. Dev, S. Kumar, S. Verma and S. Gupta,
{\it Nucl. Phys. B} {\bf 784}, 103 (2007); {\it ibid.} {\it Phys.
Rev. D} {\bf 76}, 013002 (2007).

\bibitem{qlepuni}A. Yu. Smirnov, hep-ph/0604213.

\bibitem{neeluijmpa}N. Mahajan, R. Verma and M. Gupta, {\it Int. J. Mod. Phys. A} {\bf 25}, 1 (2010).

\bibitem{neelu6zero}N. Mahajan, M. Randhawa, M. Gupta and P. S. Gill,
arXiv:1010.5640.

\bibitem{pdg}K. Nakamura {\it et al.},  {\it JPG} {\bf 37}, 075021 (2010),
updated results available at http://pdg.lbl.gov/.

\bibitem{pmns}B. Pontecorvo, {\it Zh. Eksp. Teor. Fiz. (JETP)} {\bf 33},
549 (1957); {\it ibid.} {\bf 34}, 247 (1958); {\it ibid.} {\bf
53}, 1717 (1967); Z. Maki, M. Nakagawa and S. Sakata, {\it Prog.
Theor. Phys.} {\bf 28}, 870 (1962).

\bibitem{seesaw}P. Minkowski, {\it Phys. Lett. B } {\bf 67} 421 (1977);
T. Yanagida, in {\it Proceedings of the Workshop on Unified Theory
and the Baryon Number of the Universe}, eds. O. Sawada and A.
Sugamoto (KEK, Tsukuba, 1979), p. 95; M. Gell-Mann, P. Ramond and
R. Slansky, in {\it Supergravity}, eds. F. van Nieuwenhuizen and
D. Freedman (North Holland, Amsterdam, 1979), p. 315; S.L.
Glashow, in {\it Quarks and Leptons}, eds. M. L$\rm\acute{e}$vy
{\it et al.} (Plenum, New York, 1980), p. 707; R. N. Mohapatra and
G. Senjanovic, {\it Phys. Rev. Lett.} {\bf 44}, 912 (1980).

\bibitem{xingmass}Z. Z. Xing, H. Zhang and S. Zhou, {\it Phys. Rev. D} {\bf 77}, 113016 (2008).

\bibitem{leut}H. Leutwyler, hep-ph/9602255.

\bibitem{kamland}KamLAND Collab. (A. Gando {\it et al.}), {\it Phys. Rev. D}
{\bf 83}, 052002 (2011).

\bibitem{schwetztortolavalle}T. Schwetz, M. Tortola and J. W. F.
Valle, {\it New J. Phys.} {\bf 10}, 113011 (2008).

\bibitem{fogli2649}G. L. Fogli, E. Lisi, A. Marrone, A. Palazzo
and A. M. Rotunno, {\it Phys. Rev. Lett.} {\bf 101}, 141801
(2008).

\bibitem{garcia}M. C. Gonzalez-Garcia and M. Maltoni, {\it Phys. Rept.} {\bf 460}, 1 (2008).

\bibitem{brancoperm}G.C. Branco, D. Emmanuel-Costa, R. Gonzalez Felipe and H.
Serodio, {\it Phys. Lett. B} {\bf 670}, 340 (2009).

\bibitem{neeluunpublished}N. Mahajan, {\it ``Phenomenology of quark
mixing, lepton mixing and mass matrices"}, thesis submitted to
Panjab University 2010 (unpublished).

\bibitem{frixingzhou}H. Fritzsch, Z. Z. Xing and Y. L. Zhou, {\it Phys. Lett. B}
{\bf 697}, 357 (2011).

\end{thebibliography}
\end{document}